\documentclass[12pt]{article}

\newcommand{\text}{\rm}

\begin{document}

\title{\textbf{Ghost condensates in Yang-Mills theories in nonlinear gauges } }
\author{V.E.R. Lemes, M.S. Sarandy, and S.P. Sorella \\
{\small {\textit{UERJ - Universidade do Estado do Rio de Janeiro,}}} \\
{\small {\textit{\ Rua S\~{a}o Francisco Xavier 524, 20550-013 Maracan\~{a}, 
}}} {\small {\textit{Rio de Janeiro, Brazil.}}} \and M. Picariello, \\
{\small {\textit{Universit\'{a} degli Studi di Milano, via Celoria 16,
I-20133, Milano, Italy }}}\\
{\small {\textit{and INFN\ Milano, Italy}}} \and A. R. Fazio, \\
{\small {\textit{National Research Center Demokritos, Ag. Paraskevi,
GR-153130 Athens, }}}\\
{\small {\textit{Hellenic Republic }}}}
\maketitle

\vspace{0.5cm}
{\centerline {\bf PACS: 12.38Aw, 11.15.Tk} }
\vspace{0.5cm}

\begin{abstract}
Ghost condensates of dimension two are analysed in a class of nonlinear
gauges in pure Yang-Mills theories. These condensates are related to the
breaking of the $SL(2,R)$ symmetry, present in these gauges. 
\end{abstract}

\vfill\newpage\ \makeatother

\renewcommand{\theequation}{\thesection.\arabic{equation}}

\section{Introduction}

Recently, great attention has been devoted to pure Yang-Mills theory
quantized in nonlinear gauges, the aim being that of obtaining insights
about the nonperturbative infrared behavior. For instance, the so called
Maximal Abelian gauge \cite{th,kr} has been extensively studied in the
context of the Abelian dominance hypothesis, according to which the relevant
low energy degrees of freedom for Yang-Mills should be described by an
effective abelian theory with the addition of monopoles. The condensation of
the monopoles should account for the confinement of the chromoelectric
charges, according to the dual superconductivity mechanism \cite{th,scon}.

The Maximal Abelian gauge, being a nonlinear gauge, requires the
introduction of a four ghost interaction term, needed for the
renormalizability of the model \cite{mlp,f}. As a consequence, a nontrivial
vacuum state arises \cite{ms,k,sp,dd}, corresponding to a nonvanishing value
for the ghost condensates $\left\langle f^{i\alpha \beta }c^{\alpha
}c^{\beta }\right\rangle $, $\left\langle f^{i\alpha \beta }\overline{c}%
^{\alpha }\overline{c}^{\beta }\right\rangle $, $\left\langle f^{i\alpha
\beta }c^{\alpha }\overline{c}^{\beta }\right\rangle $ and $\left\langle 
\overline{c}^{\alpha }c^{\alpha }\right\rangle $, where the index $i$ labels
the $(N-1)$ diagonal generators of the Cartan subgroup of $SU(N),$ and $%
\alpha ,\beta $ the $N(N-1)$ off-diagonal generators. These condensates turn
out to display rather interesting features. They modify the behavior of the
off-diagonal ghost propagator in the infrared region \cite{ms,k,sp} and
lower the vacuum energy density, being interpreted as a low-energy
manifestation of the trace anomaly $\left\langle T_{\mu }^{\mu
}\right\rangle $, which is related to the gluon condensate $\left\langle
\alpha F^{2}\right\rangle $.

The aim of this work is to investigate the existence of the ghost
condensates $\left\langle f^{i\alpha \beta }c^{\alpha }c^{\beta
}\right\rangle $, $\left\langle f^{i\alpha \beta }\overline{c}^{\alpha }%
\overline{c}^{\beta }\right\rangle $ in another class of nonlinear gauges 
\cite{dj} containing a ghost self-interaction term, usually referred as the
Curci-Ferrari gauge \cite{cf,ds,r}. This point could be of some help in
order to improve our understanding of the meaning of these condensates. Our
analysis shows that these condensates seem not to be related to a specific
gauge, as the Maximal Abelian gauge, being present also in the Curci-Ferrari
gauge. As pointed out in \cite{sl2r}, both the Maximal Abelian and the
Curci-Ferrari gauge display a $SL(2,R)$ symmetry whose generators act
nontrivially on the Faddeev-Popov ghosts, while leaving the gauge fields
unchanged. It turns out indeed that the ghost condensates $\left\langle
f^{i\alpha \beta }c^{\alpha }c^{\beta }\right\rangle $, $\left\langle
f^{i\alpha \beta }\overline{c}^{\alpha }\overline{c}^{\beta }\right\rangle $%
, $\left\langle f^{i\alpha \beta }c^{\alpha }\overline{c}^{\beta
}\right\rangle $ are precisely related to the dynamical breaking of $SL(2,R)$%
. It is worth mentioning here that the breaking can occur in different
channels, according to which generators are broken. More specifically, the
three generators of $SL(2,R)$, namely $\delta $, $\overline{\delta }$ and $%
\delta _{FP}\;$are known \cite{oj} to obey the algebra $\left[ \delta ,%
\overline{\delta }\right] =$ $\delta _{FP}$, where $\delta _{FP}$ denotes
the ghost number. The condensate $\left\langle f^{i\alpha \beta }c^{\alpha }%
\overline{c}^{\beta }\right\rangle $ in the Curci-Ferarri gauge has been
discussed by \cite{k1} and corresponds to the breaking of the generators $%
\delta $, $\overline{\delta }$. In the present work we shall analyse the
other condensates $\left\langle f^{i\alpha \beta }c^{\alpha }c^{\beta
}\right\rangle $, $\left\langle f^{i\alpha \beta }\overline{c}^{\alpha }%
\overline{c}^{\beta }\right\rangle $ which are related to the breaking of $%
\left( \delta ,\delta _{FP}\right) $ and of $\left( \overline{\delta }%
,\delta _{FP}\right) $, respectively. We remark also that the existence of
different channels for the ghost condensation has an analogy in
superconductivity, known as the BCS\footnote{%
Particle-particle and hole-hole pairing.} versus the Overhauser\footnote{%
Particle-hole pairing.} effect \cite{ov}. In the present case the
Faddeev-Popov charged condensates $\left\langle f^{i\alpha \beta }c^{\alpha
}c^{\beta }\right\rangle $, $\left\langle f^{i\alpha \beta }\overline{c}%
^{\alpha }\overline{c}^{\beta }\right\rangle $ would correspond to the BCS
channel, while $\left\langle f^{i\alpha \beta }c^{\alpha }\overline{c}%
^{\beta }\right\rangle $ to the Overhauser channel.

Although the analysis of the condensate $\left\langle \overline{c}^{\alpha
}c^{\alpha }\right\rangle $ is out of the aim of the present work, we
underline that, in the Maximal Abelian gauge, it is believed to be part of a
more general condensate, namely $\left( \frac{1}{2}\left\langle A_{\mu
}^{\alpha }A^{\mu \alpha }\right\rangle -\xi \left\langle \overline{c}%
^{\alpha }c^{\alpha }\right\rangle \right) $, where $\xi $ denotes the gauge
parameter. This condensate has been proposed in \cite{ope} due to its BRST\
invariance. It is expected to provide effective masses for both off-diagonal
gauge and ghost fields \cite{dd,ope}, thus playing a very important role for
the Abelian dominance. It is useful to note here that the operator $\left( 
\frac{1}{2}A^{2}-\xi \overline{c}c\right) $ generalizes to the Curci-Ferrari
gauge, displaying the important property of being multiplicatively
renormalizable \cite{mr,mr1}.

The paper is organized as follows. In Sect.2 we present the model and its
BRST quantization. Sect.3 is devoted to the evaluation of the one-loop
effective potential for the ghost condensates and to the study of the vacuum
configurations.

\section{Yang-Mills in nonlinear gauges}

Let us begin by reviewing the quantization of pure $SU(N)\;$Yang-Mills in
the Curci-Ferrari gauge. The gauge fixed action turns out to be 
\begin{equation}
S=S_{\mathrm{YM}}+S_{\mathrm{gf}}\;,  \label{scf}
\end{equation}
where $S_{\mathrm{YM}}$ is the Yang-Mills action 
\[
S_{\mathrm{YM}}=-\frac{1}{4}\int d^{4}xF^{a\mu \nu }F_{\mu \nu }^{a}\;,
\]
and $S_{\mathrm{gf}}$ denotes the nonlinear gauge fixing term with the
quartic ghost interaction 
\begin{eqnarray}
S_{\mathrm{gf}}\; &=&s\int d^{4}x\left( \overline{c}^{a}\partial A^{a}+\frac{%
\xi }{2}\overline{c}^{a}b^{a}-\frac{\xi }{4}gf^{abc}\overline{c}^{a}%
\overline{c}^{b}c^{c}\right)   \label{cfg} \\
&=&\int d^{4}x\left( b^{a}\partial _{\mu }A^{\mu a}+\frac{\xi }{2}%
b^{a}b^{a}+\,\overline{c}^{a}\partial ^{\mu }\left( D_{\mu }c\right)
^{a}\right.   \nonumber \\
&&\,\,\,\,\left. -\frac{\xi }{2}gf^{abc}b^{a}\overline{c}^{b}c^{c}-\frac{\xi 
}{8}g^{2}f^{abc}\overline{c}^{a}\overline{c}^{b}f^{cmn}c^{m}c^{n}\right) \;.
\nonumber
\end{eqnarray}
The color indices run here over all generators of $SU(N)$, \textit{i.e. }$a$,%
$b$,$c=1,....,N^{2}-1$. The operator $s$ is the nilpotent BRST operator
acting on the fields as 
\begin{eqnarray}
sA_{\mu }^{a} &=&-\left( D_{\mu }c\right) ^{a}\;,\;\;\;\;\;\;sc^{a}=\frac{g}{%
2}f^{abc}c^{b}c^{c}\;,  \label{s} \\
s\overline{c}^{a} &=&b^{a}\;\;\;\;\;\;\;,\;\;\;\;\;\;sb^{a}=0\;,  \nonumber
\end{eqnarray}
and 
\begin{eqnarray}
F_{\mu \nu }^{a} &=&\partial _{\mu }A_{\nu }^{a}-\partial _{\nu }A_{\mu
}^{a}+gf^{abc}A_{\mu }^{b}A_{\nu }^{c}\;,  \label{d} \\
\left( D_{\mu }c\right) ^{a} &=&\partial _{\mu }c^{a}+gf^{abc}A_{\mu
}^{b}c^{c}\;.  \nonumber
\end{eqnarray}
Notice that expression $\left( \ref{cfg}\right) $ contains a unique
parameter $\xi $. Moreover, as already mentioned, in addition to the BRST\
invariance, the model displays a further global $SL(2,R)$ symmetry \cite{oj}%
. The corresponding generators $\delta ,$ $\overline{\delta }$ and $\delta
_{FP}$ are given by 
\begin{eqnarray}
\delta \overline{c}^{a} &=&c^{a}\;,\;\;\;\;\;\;\;\delta b^{a}=\frac{g}{2}%
f^{abc}c^{b}c^{c}\;,  \label{delta} \\
\delta A_{\mu }^{a} &=&0\;,\;\;\;\;\;\;\;\;\delta c^{a}=0\;,  \nonumber
\end{eqnarray}
\begin{eqnarray}
\overline{\delta }c^{a} &=&\overline{c}^{a}\;,\;\;\;\;\;\;\overline{\delta }%
b^{a}=\frac{g}{2}f^{abc}\overline{c}^{b}\overline{c}^{c}\;,  \label{db} \\
\overline{\delta }A_{\mu }^{a} &=&0\;,\;\;\;\;\;\;\;\overline{\delta }%
\overline{c}^{a}=0\;,  \nonumber
\end{eqnarray}
and 
\begin{eqnarray}
\delta _{FP}c^{a} &=&c^{a}\;,\;\;\;\;\delta _{FP}\overline{c}^{a}=-\overline{%
c}^{a}\;,  \label{fp} \\
\delta _{FP}A_{\mu }^{a} &=&0\;,\;\;\;\;\;\delta _{FP}b^{a}=0\;,  \nonumber
\end{eqnarray}
with 
\begin{equation}
\left[ \delta \mathrm{{,}\overline{\delta }}\right] =\delta _{FP}\;.
\label{al}
\end{equation}
As proven in \cite{ds}, the BRST$\;$symmetry together with the $\delta $
invariance are sufficient to ensure the perturbative renormalizability of
the model, meaning that the gauge-fixing $\left( \ref{cfg}\right) $ is
stable under radiative corrections.

\section{The ghost condensates}

In order to study the ghost condensates $\left\langle f^{abc}\overline{c}^{a}%
\overline{c}^{b}\right\rangle $, $\left\langle
f^{abc}c^{b}c^{c}\right\rangle $ we first eliminate the Lagrange multiplier
field $b^{a}$. From 
\begin{equation}
\frac{\delta S}{\delta b^{a}}=\partial _{\mu }A^{\mu a}+\xi b^{a}-\frac{\xi 
}{2}gf^{abc}\overline{c}^{b}c^{c}\;,  \label{beq}
\end{equation}
we get 
\begin{eqnarray}
S &=&S_{\mathrm{YM}}+\int d^{4}x\left( -\frac{1}{2\xi }\left( \partial
A^{a}\right) ^{2}+\,\overline{c}^{a}\partial ^{\mu }\left( D_{\mu }c\right)
^{a}+\frac{g}{2}f^{abc}\partial A^{a}\overline{c}^{b}c^{c}\right) 
\label{be} \\
&&+\int d^{4}x\left( -\frac{\xi }{8}g^{2}\left( f^{abc}\overline{c}%
^{b}c^{c}\right) \left( f^{amn}\overline{c}^{m}c^{n}\right) -\frac{\xi }{8}%
g^{2}\left( f^{abc}\overline{c}^{a}\overline{c}^{b}\right) \left(
f^{cmn}c^{m}c^{n}\right) \right) \;.  \nonumber
\end{eqnarray}
Using the Jacobi identity 
\begin{equation}
\left( f^{abc}\overline{c}^{b}c^{c}\right) \left( f^{amn}\overline{c}%
^{m}c^{n}\right) =-\frac{1}{2}\left( f^{abc}\overline{c}^{a}\overline{c}%
^{b}\right) \left( f^{cmn}c^{m}c^{n}\right) \;,  \label{jac}
\end{equation}
it follows 
\begin{eqnarray}
S &=&S_{\mathrm{YM}}+\int d^{4}x\left( -\frac{1}{2\xi }\left( \partial
A^{a}\right) ^{2}+\,\overline{c}^{a}\partial ^{\mu }\left( D_{\mu }c\right)
^{a}+\frac{g}{2}f^{abc}\partial A^{a}\overline{c}^{b}c^{c}\right)   \nonumber
\\
&&-\int d^{4}x\left( \frac{\xi }{16}g^{2}\left( f^{abc}\overline{c}^{a}%
\overline{c}^{b}\right) \left( f^{cmn}c^{m}c^{n}\right) \right) \;.
\label{bef}
\end{eqnarray}
To evaluate the one-loop effective potential for the ghost condensation we
introduce auxiliary fields $\sigma ^{a},\;\overline{\sigma }^{a}\;$with
ghost number $2\,$and $-2$, so that 
\begin{equation}
-\frac{\xi }{16}g^{2}\left( f^{abc}\overline{c}^{a}\overline{c}^{b}\right)
\left( f^{cmn}c^{m}c^{n}\right) \;\Rightarrow \mathcal{L}_{\sigma \overline{%
\sigma }}  \label{hbs}
\end{equation}
where 
\begin{equation}
\mathcal{L}_{\sigma \overline{\sigma }}=-\frac{1}{\xi g^{2}}\sigma ^{a}%
\overline{\sigma }^{a}\,+\frac{\overline{\sigma }^{a}}{4}f^{abc}c^{b}c^{c}-%
\frac{\sigma ^{a}}{4}f^{abc}\overline{c}^{b}\overline{c}^{c}\;\;.
\label{cfls}
\end{equation}
The relevant part of the action for the evaluation of the one-loop effective
potential is given by 
\begin{eqnarray}
S_{c\overline{c}}^{\mathrm{quad}} &=&\int d^{4}x\,\left( -\frac{1}{\xi g^{2}}%
\sigma ^{a}\overline{\sigma }^{a}\;+\overline{c}^{a}\partial ^{2}c^{a}+\frac{%
\overline{\sigma }^{a}}{4}f^{abc}c^{b}c^{c}-\frac{\sigma ^{a}}{4}f^{abc}%
\overline{c}^{b}\overline{c}^{c}\right) \;  \nonumber \\
&=&\int d^{4}x\left[ -\frac{1}{\xi g^{2}}\sigma ^{a}\overline{\sigma }^{a}+%
\frac{1}{2}\left( 
\begin{tabular}{ll}
$\overline{c}^{a}$ & $c^{a}$%
\end{tabular}
\right) \mathcal{M}^{ab}\left( 
\begin{tabular}{l}
$\overline{c}^{b}$ \\ 
$c^{b}$%
\end{tabular}
\right) \right] \;,  \label{seffcf}
\end{eqnarray}
where $\mathcal{M}^{ab}$ denotes the $\left( N^{2}-1\right) \times \left(
N^{2}-1\right) $ matrix 
\begin{equation}
\mathcal{M}^{ab}=\left( 
\begin{tabular}{ll}
$-\frac{1}{2}\sigma ^{c}f^{cab}$ & $\partial ^{2}\delta ^{ab}$ \\ 
$-\partial ^{2}\delta ^{ab}$ & $\frac{1}{2}\overline{\sigma }^{c}f^{cab}$%
\end{tabular}
\right) \;.  \label{mabcf}
\end{equation}
For the one-loop effective potential we get 
\begin{equation}
V^{\mathrm{eff}}(\sigma ,\overline{\sigma })=\frac{1}{\xi g^{2}}\sigma ^{a}%
\overline{\sigma }^{a}+\frac{i}{2}\mathrm{tr\log \det }\mathcal{M}^{ab}.
\label{vsscf}
\end{equation}
where $\sigma ^{a},\overline{\sigma }^{a}$ have to be considered constant
fields and where we have factorized the space-time volume.

Let us proceed by working out in detail the example of $SU(2)$. In this
case, we have $f^{abc}=\varepsilon ^{abc}\;(\varepsilon ^{123}=1)$, and $%
\mathcal{M}^{ab}$ is a $6\times 6$ matrix. After a simple computation one
has 
\begin{equation}
V^{\mathrm{eff}}(\sigma ,\overline{\sigma })=\frac{1}{\xi g^{2}}\sigma ^{a}%
\overline{\sigma }^{a}+i\int \frac{d^{4}k}{\left( 2\pi \right) ^{4}}\log
\left( \left( k^{2}\right) ^{2}+\frac{\sigma ^{a}\overline{\sigma }^{a}}{4}%
\right) \;.  \label{vss1cf}
\end{equation}
From 
\begin{equation}
\int \frac{d^{4}k}{\left( 2\pi \right) ^{4}}\log \left( \left( k^{2}\right)
^{2}+\varphi ^{2}\right) =\frac{i}{32\pi ^{2}}\varphi ^{2}\left( -\frac{1}{%
\varepsilon }-2\gamma +2\log 4\pi -\log \frac{\varphi ^{2}}{\mu ^{4}}%
+3\right) \;,  \label{ighcf}
\end{equation}
and making use of the minimal subtraction scheme, the effective potential $%
V^{\mathrm{eff}}(\sigma ,\overline{\sigma })$ is found to be 
\begin{equation}
V^{\mathrm{eff}}(\sigma ,\overline{\sigma })=\frac{1}{\xi g^{2}}\sigma ^{a}%
\overline{\sigma }^{a}+\frac{1}{32\pi ^{2}}\frac{\sigma ^{a}\overline{\sigma 
}^{a}}{4}\left( \log \frac{\sigma ^{a}\overline{\sigma }^{a}}{4\left( 4\pi
\right) ^{2}\mu ^{4}}-3+2\gamma \right) \;.  \label{veffms}
\end{equation}
Let us look now at the minimum of the potential $\left( \ref{veffms}\right) .
$ It is given by the condition 
\begin{equation}
\sigma _{\mathrm{\min }}^{a}\overline{\sigma }_{\mathrm{\min }}^{a}=64\pi
^{2}\mu ^{4}e^{\left( 2-2\gamma \right) }\exp \left( -\frac{128}{\xi }\frac{%
\pi ^{2}}{g^{2}}\right) \;,  \label{min}
\end{equation}
which is physically consistent for any $\xi >0.\;$Setting now 
\begin{equation}
\sigma ^{a}=\sigma \delta ^{a3}\;,\;\;\;\;\overline{\sigma }^{a}=\overline{%
\sigma }\delta ^{a3}\;\;,  \label{3d}
\end{equation}
for the vacuum configuration we obtain 
\begin{equation}
\sigma _{\mathrm{\min }}\overline{\sigma }_{\mathrm{\min }}=64\pi ^{2}\mu
^{4}e^{\left( 2-2\gamma \right) }\exp \left( -\frac{128}{\xi }\frac{\pi ^{2}%
}{g^{2}}\right) \;.  \label{bcff}
\end{equation}
The nontrivial minimum configuration $\left( \ref{bcff}\right) $ means that
the local operators $\varepsilon ^{3bc}c^{b}c^{c},$ $\varepsilon ^{3bc}%
\overline{c}^{b}\overline{c}^{c}$ acquire a nonvanishing vacuum expectation
value, implying a dynamical breaking of $SL(2,R)$. Indeed 
\begin{eqnarray}
\left\langle \varepsilon ^{3bc}c^{b}c^{c}\right\rangle  &=&\frac{1}{2}
\left\langle \delta _{FP}\left( \varepsilon ^{3bc}c^{b}c^{c}\right)
\right\rangle =\left\langle \delta \left( \varepsilon ^{3bc}\overline{c}%
^{b}c^{c}\right) \right\rangle \;,  \label{br} \\
\left\langle \varepsilon ^{3bc}\overline{c}^{b}\overline{c}^{c}\right\rangle
&=&-\frac{1}{2}\left\langle \delta _{FP}\left( \varepsilon ^{3bc}\overline{c}%
^{b}\overline{c}^{c}\right) \right\rangle =\left\langle \overline{\delta }%
\left( \varepsilon ^{3bc}c^{b}\overline{c}^{c}\right) \right\rangle \;. 
\nonumber
\end{eqnarray}
Observe also that the vacuum configuration $\left( \ref{3d}\right) $, $%
\left( \ref{bcff}\right) $ leaves invariant the Cartan subgroup $U(1)$ of $%
SU(2),$ in analogy with the Maximal Abelian gauge.

It remains now to face the important question of the stability of the
nontrivial vacuum $\left( \ref{bcff}\right) $, whose physical meaning relies
on the existence of a positive value for the parameter $\xi $, which plays
the role of a coupling constant for the quartic ghost self-interaction. A
natural choice for the parameter $\xi $ would be given by the fixed point of
the corresponding renormalization group function $\beta _{\xi }$. This would
require the knowledge of the existence of a nonperturbative fixed point for $%
\xi $, which is beyond our present possibilities. However, it worth to
remind that a detailed analysis of the renormalization of the Curci-Ferrari
gauge has been performed at one-loop level by \cite{mr}. Recently, the
authors \cite{mr1} have worked out the two and three loops contributions. In
particular, according to \cite{mr,mr1}, the running of $\xi $ at one-loop
order is found to be 
\begin{equation}
\beta _{\xi }=\frac{\mu }{\xi }\frac{\partial \xi }{\partial \mu }=\left( 
\frac{13}{3}-\frac{\xi }{2}\right) \frac{g^{2}N}{16\pi ^{2}}\;,  \label{ba}
\end{equation}
showing that the value $\xi =26/3$ is a fixed point, matching the
requirements for a nontrivial vacuum. Although one cannot provide a
definitive answer, these results give evidences for the existence of the
ghost condensation in the Curci-Ferrari gauge.

\section*{Acknowledgments}

The Conselho Nacional de Desenvolvimento Cient\'{\i }fico e Tecnol\'{o}gico
CNPq-Brazil, the Funda{\c{c}}{\~{a}}o de Amparo {\ a Pesquisa do Estado do
Rio de Janeiro (Faperj), the SR2-UERJ and the MIUR-Italy are acknowledged
for the financial support. M. Picariello is grateful to the Theoretical
Physics Department of UERJ for kind hospitality. }The work of A.R. Fazio was
supported by EEC Grant no. HPRN-CT-1999-00161 and partially by Fondazione
''Angelo Della Riccia''-Ente morale.

\end{document}